\documentclass[reprint,superscriptaddress,amsmath,amssymb,longbibliography]{revtex4-1}

\usepackage{graphicx, nicefrac}
\usepackage{array}
\usepackage{amsmath,amsfonts,amssymb}
\usepackage[ansinew]{inputenc}
\usepackage{color, soul}
\usepackage{calc}

\newcommand{\Figref}[1]{Fig.~\ref{#1}}

\newcommand{\ignore}[1]{}

\begin{document}
\title{Gating single-molecule fluorescence with electrons}

\author{Katharina Kaiser}
\email{katharina.kaiser@uni-goettingen.de}
\affiliation{Universit\'e de Strasbourg, CNRS, IPCMS, UMR 7504, F-67000 Strasbourg, France}
\affiliation{IV. Physical Institute -- Solids and Nanostructures,
Georg-August-Universit\"{a}t G\"{o}ttingen, G\"{o}ttingen, 37077, Germany}

\author{Michelangelo Romeo}
\affiliation{Universit\'e de Strasbourg, CNRS, IPCMS, UMR 7504, F-67000 Strasbourg, France}

\author{Fabrice Scheurer}
\affiliation{Universit\'e de Strasbourg, CNRS, IPCMS, UMR 7504, F-67000 Strasbourg, France}

\author{Guillaume Schull}
\email{guillaume.schull@ipcms.unistra.fr}
\affiliation{Universit\'e de Strasbourg, CNRS, IPCMS, UMR 7504, F-67000 Strasbourg, France}

\author{Anna Ros\l awska}
\email{a.roslawska@fkf.mpg.de}
\affiliation{Universit\'e de Strasbourg, CNRS, IPCMS, UMR 7504, F-67000 Strasbourg, France}
\affiliation{Max Planck Institute for Solid State Research, Stuttgart, 70569, Germany}

\begin{abstract}
   
Tip-enhanced photoluminescence (TEPL) measurements are performed with sub-nanometer spatial resolution on individual molecules decoupled from a metallic substrate by a thin NaCl layer. TEPL spectra reveal progressive fluorescence quenching with decreasing tip-molecule distance when electrons tunneling from the tip of a scanning tunneling microscope are injected at resonance with the molecular states. Rate equations based on a many-body model reveal that the luminescence quenching is due to a progressive population inversion between the ground neutral (S$_0$) and the ground charge ($D_0^-$) states of the molecule occurring when the current is raised. We demonstrate that both the bias voltage and the atomic-scale lateral position of the tip can be used to gate the molecular emission. Our approach can in principle be applied to any molecular system, providing unprecedented control over the fluorescence of a single molecule. 
       
\end{abstract}

\date{\today}

\maketitle

Control over the optical properties of individual molecules can be achieved by modifying their chemical structure or influencing their local environment. In that respect, the ability to switch the molecule from a bright to a dark state is one of the most striking modifications of these characteristics. Such a mechanism is employed in, for example, superresolution microscopy to localize individual emitters\cite{Folling2008,Heilemann2009, Vogelsang2009} and holds the potential to be applied in sensors\cite{Brasselet2000} or even molecular quantum devices\cite{Toninelli2021} to gate the properties of single-photon emitters\cite{Yu2022}. Usually, switching the chromophore to a dark state relies on a transition to a long-lived and non-emissive triplet state or on modifying the redox state of the molecule by attaching positive or negative charges to prevent the formation of charge-neutral excited states\cite{Brasselet2000,Ji2015a, Vogelsang2009, Dickson1997}. The latter process can be induced on demand by an electric stimulus, as is the case in macroscopic electrochromic materials\cite{Ji2015a, Gu2022}. However, gating the photoluminescence of a targeted molecule by the controlled injection of individual charges remains a thought experiment.

Scanning tunneling microscopy (STM) allows the controlled charge transfer to single molecules with angstrom-scale precision\cite{Repp2005,Nazin2005a,Swart2011,Fernandez-Torrente2012,Hauptmann2013,Cochrane2018, Kaiser2023a}. At the same time, the field enhancement in the junction between the metallic sample and the atomically sharp metallic tip, a geometry denoted as plasmonic picocavity\cite{Benz2016, Roslawska2022}, enables probing electrically-driven STM-induced luminescence (STML)\cite{Qiu2003, Imada2016, Zhang2016, Doppagne2017, Luo2019, Cao2021, Doppagne2018, Rai2020, Dolezal2021a, Dolezal2021b, Dolezal2022,Jiang2023, Hung2023, Roslawska2022, Rai2023, Kaiser2024} or optically-driven tip-enhanced photoluminescence (TEPL)\cite{Yang2020,Imada2021, Imai-Imada2022, Roslawska2023, Dolezal2024} of individual molecules adsorbed on ultrathin insulating films with nearly atomic precision. STML has, in a few cases, been used to address the fluorescence of charged species, however, the electrical excitation renders it impossible to separate luminescence and charge transport from each other and to control them separately\cite{Doppagne2018, Rai2020, Dolezal2021a, Dolezal2021b, Dolezal2022,Jiang2023, Hung2023, Rai2023, Kaiser2024}.

Here, we show that we can deliberately manipulate the photoluminescence yield from a single molecule located in the double barrier tunneling junction of an STM, up to nearly full luminescence quenching, by resonant charge transport through the molecule. We achieve this by adjusting the tip-molecule distance, the applied bias voltage, and the in-plane position of the tip with respect to the molecule. This versatile control of the fluorescence yield of a single molecule with sub-nm precision is explained using a straightforward many-body description of our system.

The experiments were performed with a low-temperature (6 K) ultrahigh vacuum STM (Unisoku USM1400) equipped with two adjustable lenses used to focus the beam of a laser to and collect the luminescence from the junction of the STM (more details in Supplementary Materials). We used an Ag(111) substrate that was cleaned by successive argon-ion sputtering and annealing cycles. NaCl was thermally sublimed onto Ag(111) kept at room temperature and then mildly annealed to obtain two to four-monolayer thick NaCl films. The molecules were thermally sublimed onto the cold (6 K) NaCl/Ag(111) substrate. We used electrochemically etched Ag tips, that were sputtered and annealed in the ultrahigh vacuum. The plasmonic response of the tips was optimized by voltage pulses and gentle indentations into the Ag(111) substrate.

\begin{figure}[!ht]
    \includegraphics[width = \linewidth]{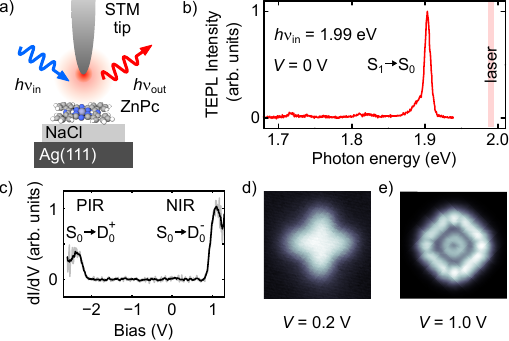}
    \caption{a) Sketch of the experiment.  b) TEPL spectrum recorded on ZnPc. $h\nu_{\rm in} =$ 1.99 eV, laser power $P$ = 13 $\mu$W, integration time $t$ = 10 s, $V$ = 0 V. c) d$I/$d$V$ spectrum recorded on ZnPc. Set-point: $V$ = 1.3 V, $I$ = 20 pA. d) Constant-current STM image recorded on 3 ML NaCl, $V$ = 0.2 V, $I$ = 3 pA. e) Constant-height STM image recorded on 4 ML NaCl, $V$ = 1 V. Both images are 3 $\times$ 3 nm$^2$.}
    \label{Fig1}
\end{figure}
\noindent

\Figref{Fig1}a shows a scheme of the experiment. A zinc phthalocyanine (ZnPc) molecule deposited on 4 monolayer (ML) thick NaCl is optically excited by a tunable laser source focused on the tip-sample junction. Within the junction, the electromagnetic field is strongly amplified between the tip apex and the metallic substrate acting as a plasmonic picocavity. This leads to increased absorption and a strong enhancement of radiative transitions (i.e., Purcell effect) so that the TEPL signal of individual molecules can be detected in the far field\cite{Yang2020, Imada2021, Imai-Imada2022, Roslawska2023, Dolezal2024}. A typical single-molecule TEPL spectrum obtained with an incident photon energy of $h\nu_{\rm in} =$ 1.99 eV is shown in \mbox{\Figref{Fig1}b}. It reveals a prominent emission line at an energy $h\nu_{\rm out} =$ 1.91 eV, and low-energy vibronic peaks. Both stem from the S$_{1}$ $\xrightarrow{}$ S$_{0}$ transition of neutral ZnPc\cite{Murray2011, Zhang2016}. \Figref{Fig1}c shows a differential conductance (d$I/$d$V$) spectrum of ZnPc/4 ML NaCl/Ag(111) with two distinct features at $V$ = -2.1 V and $V$ = 0.8 V, corresponding to the positive and negative ion resonance (PIR and NIR), respectively\cite{Repp2005}. For voltages between the PIR and the NIR (referred to as in-gap conditions), the charge transport through the double-barrier tunneling junction is non-resonant and only perturbed by the presence of the molecule \cite{Grewal2023}. In contrast, at the PIR and NIR, the charge transport through the junction is dominated by a two-step process where the molecule is first charged by tunneling from the tip and then discharged by tunneling to the sample. Thus, for resonant conditions, the molecule repetitively switches between a neutral and a charged configuration. \Figref{Fig1}d and \Figref{Fig1}e show STM images of ZnPc recorded in these two transport regimes, namely for in-gap ($V$ = 0.2 V) and at NIR ($V$ = 1 V), respectively, that is \Figref{Fig1}e shows the density of the lowest unoccupied molecular orbital (LUMO). Note that the image in \Figref{Fig1}d was recorded in constant current mode for a molecule adsorbed on 3 ML NaCl, other data shown in the main manuscript are recorded on 4 ML.

\begin{figure*}[!ht]
    \includegraphics[width = 0.75\linewidth]{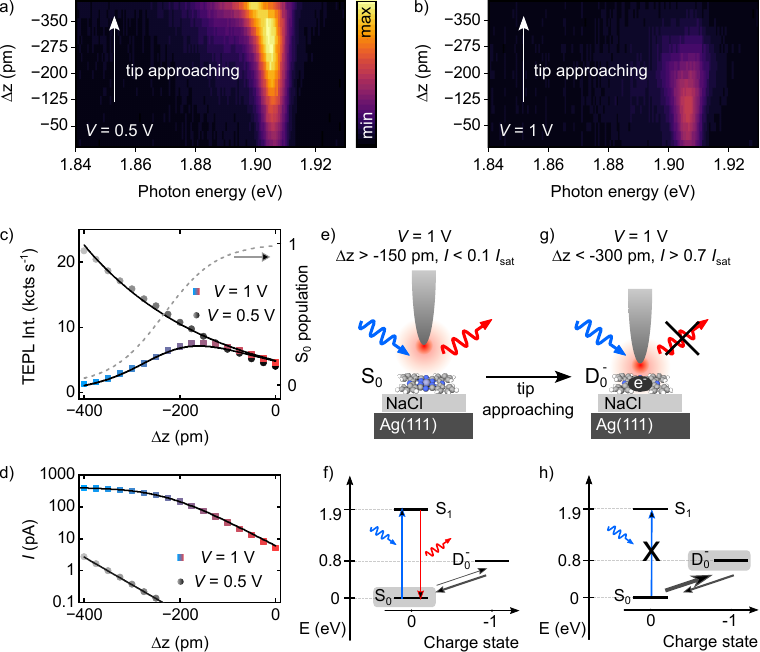}
    \caption{a,b) Two-dimensional representation of successive TEPL spectra recorded as a function of the relative tip-sample distance for $V$ = 0.5 V (a) and $V$ = 1 V (b). The bottom spectra are recorded at $\Delta z$ = 0, then the tip is approached in 25 pm intervals. $\Delta z$ = 0 is defined for a setpoint of $V$ = 1 V, $I$ = 5 pA. $h\nu_{\rm in} =$ 1.99 eV, $P$ = 3.5 $\mu$W, $t$ = 4 s. The color scale is the same for both panels. c) Integrated TEPL intensity (1.88 eV $< h\nu_{out} <$ 1.92 eV) for $V$ = 0.5 V (grey-scale circles, panel a) and $V$ = 1 V (red-blue squares, panel b) as a function of $\Delta z$. The black and red shading indicates the $\Delta z$ range in which the tunnel current increases exponentially, light-grey and blue shading indicate the saturation regime. The solid lines are fits to the model discussed in the main text, the dashed grey line shows the relative population of the ground S$_0$ state. d) Tunneling current recorded together with measurements in c), the solid line for the $V$ = 1 V data is a fit to Eq. 1, and the solid line for the $V$ = 0.5 V is an exponential fit. e-h) Sketches of the experimental conditions (e, g) and the many-body diagrams for the large (f) and small (h) relative tip-sample distances. The grey boxes in f,h) indicate the state that is populated most of the time.}
    \label{Fig2}
\end{figure*}
\noindent 

In the following, we investigate how charge transport in the two regimes influences the optical properties of the neutral molecule. \Figref{Fig2}a shows a series of TEPL spectra recorded in-gap ($V$ = 0.5 V) for different relative tip-molecule distances $\Delta z$. This parameter is defined relatively to an initial tip-molecule separation ($\Delta$z = 0) at a setpoint of $V$ = 1 V and $I$ = 5 pA. Upon approaching the tip to the molecule (decreasing $\Delta z$), the $S_1\xrightarrow{}S_0$ peak in TEPL increases in intensity, broadens from 6 meV to more than 10 meV (full width at half maximum), as extracted from Lorentzian fits, and shifts by 6 meV towards smaller energies at very small distances. This is in line with previous experiments\cite{Roslawska2022}, which assigned a similar peak broadening and increase in emission intensity to an increased coupling between the molecules' transition dipole moments and the picocavity plasmons upon reduction of the relative tip-molecule distance. The observed red-shift can be attributed to a combination of Lamb and Stark effects. The same $\Delta z$ dependence, but recorded at the NIR ($V$ = 1 V), is shown in \Figref{Fig2}b. Here, as well, we observe a broadening of the peak in TEPL for decreasing $\Delta z$. For $\Delta$z $<$ -150 pm, however, the photoluminescence intensity decreases and is almost entirely suppressed for $\Delta$z $<$ -400 pm.

This gradual reduction of the emission yield with decreasing $\Delta z$, which is only observed at resonance, suggests a quenching mechanism related to the transient charging of the molecule. To confirm this, we recorded the tunnel current simultaneously with the integrated TEPL intensity of the $S_1\xrightarrow{}S_0$ emission line (\Figref{Fig2}c, d). Note that, for $V$ = 0.5 V and $\Delta z > -250\,\text{pm}$, the current is too low to be measured. At $\Delta z > -150$ pm, where $I(\Delta z)$ recorded at resonance follows the expected exponential dependence (red squares in \Figref{Fig2}d), the TEPL intensity increases similarly for both bias voltages. Upon approaching the molecule ($\Delta z$ $<$ -150 pm), the $I(\Delta z)$ curve measured at resonance deviates from the exponential dependence, while the TEPL intensity slowly decreases. This behaviour contrasts with the TEPL intensity recorded in-gap that increases monotonously with the tip approach (gray-scale circles in \Figref{Fig2}c). Eventually, at very small relative tip-molecule distances and for resonant tunneling conditions (blue squares in \Figref{Fig2}c, d), the current saturates ($I_{sat} \approx $ 400 pA), and the TEPL intensity is nearly fully quenched.

The saturation in the tunnel current is a direct consequence of the charge transport mechanism that takes place in the double-barrier tunneling junction of our experiment\cite{Steurer2014,Kaiser2023a}. At NIR, the molecule is successively (i) charged negatively by an electron tunneling through the vacuum barrier from the tip to the molecule, and then (ii) discharged by a second tunneling event taking place through the NaCl layer. The overall tunnel current is composed of these two consecutive events and is limited by the process with the lower tunneling probability, \textit{i.e.}, the slower process. While the tunneling probability through the vacuum gap varies with $\Delta z$, the tunneling probability through the NaCl is merely determined by the thickness of the NaCl layer and thus remains constant as the tip approaches. At large relative tip-molecule distances, tunneling through the vacuum is, therefore, the rate-limiting process  ($\Delta z >$ -150 pm, \Figref{Fig2}e, f), and the tunneling current increases exponentially with decreasing $\Delta z$. At smaller relative tip-molecule distances ($\Delta z <$ -150 pm) the progressive saturation of the current indicates that the discharging process through the NaCl becomes the rate-limiting step. This affects the relative populations of the neutral $S_0$ and charged $D_0^-$ states of the molecule, as indicated in the many-body diagrams in \Figref{Fig2}f, h: At large relative tip-molecule distances ($\Delta z >$ -150 pm), the charging rate indicated by the dark grey arrow is low and the molecule spends most of the time in the neutral state $S_0$. Optical excitation of the S$_{0}$ $\xrightarrow{}$ S$_{1}$ transition by incoming photons and the emission from the reverse S$_{1}$ $\xrightarrow{}$ S$_{0}$ transition (\Figref{Fig2}f) is thus possible. For decreasing $\Delta z$, the population of the negatively charged $D_0^-$ state increases at the expense of the $S_0$ population, up to a predominant occupation of $D_0^-$ in the saturation regime (\Figref{Fig2}g, h). Since the D$_{0}^{-}$ $\xrightarrow{}$ S$_{1}$ transition cannot be induced by a photon, which carries a net zero charge, the formation of a neutral exciton is gradually hindered and the TEPL signal vanishes. Thus, the controllable transition to a configuration where the charged doublet state is occupied most of the time, instead of the neural singlet, is responsible for the quenching of the luminescence, akin to the process taking place in electrochromic systems. A similar effect is expected to take place at PIR when the D$_{0}^{+}$ population is promoted. At this condition, however, electrical excitation of S$_{1}$ via D$_{0}^{+}$ leads to a strong STML signal at the same photon energy\cite{Dolezal2019, Dolezal2021a, Hung2023}, precluding the observation of the photoluminescence quenching.

To quantify the mechanism behind the photoluminescence quenching, we develop a model encompassing the many-body transitions between the relevant states in our system, $S_0$, $S_1$ and $D_0^-$, illustrated in \Figref{Fig2}f, h (see Supplementary Materials for more details). The total tunnel current through the double-barrier tunneling junction can be expressed using the tunneling rates through vacuum, $\Gamma_{vac}(\Delta z)$ ($S_0$ $\xrightarrow{}$ $D_0^-$ transition), and NaCl, $\Gamma_{d}$ ($D_0^-$ $\xrightarrow{}$ $S_0$ transition), respectively\cite{Kaiser2023a}:

\begin{equation}
    I(\Delta z) = \frac{q\Gamma_{vac}^{0}}{\rm{exp(2\kappa_{vac}\Delta z)}+\Gamma_{vac}^{0} \tau_d}
    \label{Eq1}
\end{equation}
with the elementary charge $q$, the tunneling rate from the tip $\Gamma_{vac}^{0}$ for $\Delta z$ = 0, the decay constant $\kappa_{vac}$ and the $D_0^-$ lifetime $\tau_d$. The steady-state ground state population $N_{S_0}$ is then given by:
\begin{equation}
    N_{S_0}=\frac{1}{1+\tau_d\Gamma_{vac}^{0}\rm{exp(-2\kappa_{vac}\Delta z)}}
    \label{eq:pop}
\end{equation}
Fitting $I(\Delta z)$ recorded for $V=1$\,V in \Figref{Fig2}d (solid line) with the expression in Eq. \eqref{Eq1} yields $\tau_d$ $\approx$ 400 ps, in line with recent reports on decoupled emitters \cite{Kaiser2023a, Roslawska2018}. Once the tunneling rate from the tip becomes comparable to the tunneling rate from the substrate, which in our system is observed for $\Delta z$ $<$ -150 pm, $N_{S_0}$ begins to substantially deviate from 1 (dashed line in \Figref{Fig2}c).

Next we analyse the $S_0$ $\xrightarrow{}$ $S_1$ and $S_1$ $\xrightarrow{}$ $S_0$ transitions that are responsible for the observed photoluminescence. The total light intensity can be written as $P(\Delta z) = \Gamma_{ex}N_{S_0}$, where $\Gamma_{ex}$ is the excitation rate. In agreement with recent literature\cite{Yang2020, Imai-Imada2022}, we find that $\Gamma_{ex}$($\Delta z$) is well-described by an exponential function, which reflects the strongly increased plasmonic field when the tip approaches the molecule. Therefore
\begin{equation}
    P(\Delta z) = \Gamma_{ex}^{0}\rm{exp(-\alpha \Delta z)}N_{S_0}
    \label{eq:int}
\end{equation}
 with $\Gamma_{ex}^{0}$ being the excitation rate for $\Delta z$ = 0 and $\alpha$ the decay constant. Fitting the data in \Figref{Fig2}c with Eq. (\ref{eq:int}) and $N_{S_0}$ = 1 for $V$ = 0.5 V, that is assuming very short excited state lifetime \cite{Roslawska2022, Dolezal2024} and thus $N_{S_1} \approx$ 0, we obtain a very good agreement with the experiment. For fitting the curve recorded at $V$ = 1 V, we use the same values of the parameters ($\Gamma_{ex}^{0}$ and $\alpha$) but $N_{S_0}$ described by Eq. (\ref{eq:pop}). Again, our model is in excellent agreement with the experimental data. Comparing the $P(\Delta z, V $= 1 V$)$ and $N_{S_0}(\Delta z)$ curves in \Figref{Fig2}c, we find that the quenching effect onsets already for $N_{S_0} \approx$ 0.8, when $\Delta z =$ -150 pm.

\begin{figure}[!ht]
    \includegraphics[width = \linewidth]{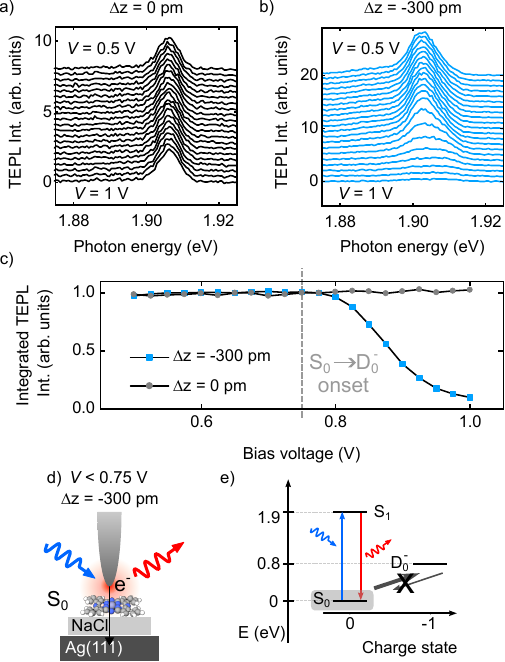}
    \caption{a,b) TEPL spectra recorded as a function of bias voltage for two different $\Delta z$. The voltage increases from the top to the bottom spectra in 25 mV intervals. $h\nu_{\rm in} =$ 1.99 eV, $P$ = 3.5 $\mu$W, $t$ = 5 s. $\Delta z$ = 0 is defined for $V$ = 1 V, $I$ = 5 pA. c) Integrated and normalized TEPL intensity from the panels a) and b), integration range: 1.88 eV $< h\nu_{out} <$ 1.92 eV. d) Sketch of the experimental conditions for $V$ $<$ 0.75 V and $\Delta z$ = -300 pm. Note that the tunneling current does not populate the D$_{0}^{-}$ state. e) Many-body diagram corresponding to the configuration in d).}
    \label{Fig3}
\end{figure}
\noindent

Since the ability to excite a molecule with incoming photons depends on the relative population of $S_0$, we can conveniently gate the TEPL yield by the applied bias voltage. \mbox{\Figref{Fig3}a} and \mbox{\Figref{Fig3}b} show constant-height TEPL spectra as a function of the applied bias voltage for large ($\Delta z = 0$\, pm) and small ($\Delta z = -300$\, pm) relative tip-molecule distances, respectively. At $\Delta z = 0$\, pm (\mbox{\Figref{Fig3}a}), the TEPL spectra remain unaffected by the change in bias voltage from 0.5 V to 1 V. In contrast, at small relative tip-molecule distances (\Figref{Fig3}b), increasing the bias from 0.5 V to 1 V results in gradual quenching of the TEPL signal. The integrated intensity as a function of the bias voltage for these two distances is compared in \Figref{Fig3}c. At large relative tip-molecule distances (gray circles), the TEPL signal intensity remains constant. For small distances (blue squares), we find that the TEPL intensity begins decreasing at $V > 0.75$\, V, a value that agrees very well with the onset of the NIR (\Figref{Fig1}d). At large distances, the tunneling probability between the tip and molecule is low, and thus the relative population of S$_{0}$ remains high regardless of the applied voltage. This is also true at small distances to the molecule for low bias voltages, as the D$_{0}^{-}$ state cannot be accessed. Hence, in both conditions, the TEPL signal remains intense (\Figref{Fig3}d, e). In contrast, at small distances and at bias voltages that are resonant with the NIR of the molecule, the charged state is efficiently populated, and the luminescence is quenched. Overall, this shows that one can gate the photoluminescence of a single molecule by tuning the voltage applied across the junction and the distance to one of the electrodes.

\begin{figure}[!ht]
    \includegraphics[width = \linewidth]{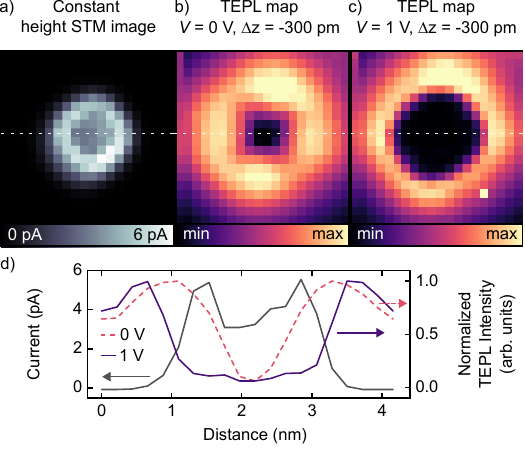}
    \caption{a) Constant height STM image of ZnPc, $V$ = 1 V. b, c) TEPL maps on the same molecule as in a), the parameters are indicated in the figure, $h\nu_{\rm in} =$ 1.99 eV, integration time $t$ = 2 s per pixel. Integration range: 1.88 eV $< h\nu_{out} <$ 1.92 eV. The size of all images is 4.4 $\times$ 4.4 nm$^2$. d) Cross-section along the white dashed lines in a-c). Current is plotted on the left axis and normalized TEPL intensity on the right axis.}
    \label{Fig4}
\end{figure}
\noindent

The local character of this TEPL gating is demonstrated by the TEPL intensity dependence on the tip position. \Figref{Fig4}a shows the spatial extent of the electronic orbital at $V=1$\,V (LUMO). \mbox{\Figref{Fig4}b, c} show maps of the TEPL intensity recorded atop the molecule in-gap (b) and at NIR (c) at a constant height of $\Delta z = -300$ pm.  The TEPL map recorded at $V$ = 0 V reveals a radially symmetrical pattern and a low TEPL intensity in the middle of the molecule. This expected spatial dependence originates from the coupling of the picocavity plasmon to the doubly-degenerate transition dipoles of S$_1$ for ZnPc. The low intensity in the center arises due to the symmetry of the system which leads to a vanishing coupling between the picocavity field and the molecular transition density\cite{Neuman2018, Yang2020}. At resonant tunneling conditions ($V$ = 1 V, \Figref{Fig4}c), the TEPL map reveals a much broader dark center compared to the map recorded at $V$ = 0 V. This dark region matches well the spatial extent of the LUMO in the constant height current map in \Figref{Fig4}a. In contrast, in regions where resonant tunneling through the molecule is not possible (at tip positions exceeding the lateral extent of the LUMO) the intensity pattern of \Figref{Fig4}b is recovered. This demonstrates that, for $V$ = 1 V and $\Delta z = -300$ pm, as soon as direct tunneling into the ZnPc is possible, we can deliberately quench the TEPL yield by controllably decreasing the relative population of $S_0$. This is further illustrated in the cross-section in \Figref{Fig4}d where the onset of the orbital (increase of the current) coincides with the onset of TEPL quenching. The $S_0$ occupancy in the current saturation regime (\Figref{Fig2}d) is entirely governed by the substrate-mediated discharging rate and is therefore independent of the tip position. Overall, this photoluminescence quenching strategy can be generalized to other molecules as we demonstrate in the Supplementary Materials for a free-base phthalocyanine, H$_2$Pc.

In conclusion, we have demonstrated that we can gate the photoluminescence of a single molecule with sub-nm precision.
We use the tip of an STM simultaneously as an atomically precise electrode, which allows us to modify the charge state of a molecule, and as an optical antenna to probe the chromophore's photoluminescence. By selectively injecting individual charges into the molecule we can controllably and gradually shift the time-averaged population of the molecule from its neutral $S_0$ ground state to a transient charge state $D_0^-$ from which the neutral exciton cannot be formed. Our approach can be used, for example, to selectively alter the properties of individual chromophores in a molecular layer or assembly. This can influence for example coherent dipole-dipole coupling\cite{Zhang2016,Doppagne2017, Luo2019} or energy transfer mechanisms\cite{Imada2016,Cao2021}, thus modifying the collective optical responses of the system.
\\

We thank Tom{\'a}\v{s} Neuman and St\'ephane Berciaud for fruitful discussions and Virginie Speisser and Halit Sumar for technical support. This project has received funding from the European Research Council (ERC) under the European Union's Horizon 2020 research and innovation program (grant agreement No 771850). This work of the Interdisciplinary Thematic Institute QMat, as part of the ITI 2021 2028 program of the University of Strasbourg, CNRS and Inserm, was supported by IdEx Unistra (ANR 10 IDEX 0002), and by SFRI STRAT'US project (ANR 20 SFRI 0012) and EUR QMAT ANR-17-EURE-0024 under the framework of the French Investments for the Future Program. K. K. acknowledges funding from the SNSF under the Postdoc.Mobility grant agreement No 206912. A. R. acknowledges funding from the Emmy Noether Programme of the Deutsche Forschungsgemeinschaft (DFG, German Research Foundation) - 534367924.

\bibliographystyle{naturemag}

\bibliography{chargingTEPL}

\end{document}


\title{Supplementary information for\\
Gating single-molecule fluorescence with electrons}

\author{Katharina Kaiser}
\email{katharina.kaiser@uni-goettingen.de}
\affiliation{Universit\'e de Strasbourg, CNRS, IPCMS, UMR 7504, F-67000 Strasbourg, France}
\affiliation{IV. Physical Institute -- Solids and Nanostructures,
Georg-August-Universit\"{a}t G\"{o}ttingen, G\"{o}ttingen, 37077, Germany}

\author{Michelangelo Romeo}
\affiliation{Universit\'e de Strasbourg, CNRS, IPCMS, UMR 7504, F-67000 Strasbourg, France}

\author{Fabrice Scheurer}
\affiliation{Universit\'e de Strasbourg, CNRS, IPCMS, UMR 7504, F-67000 Strasbourg, France}

\author{Guillaume Schull}
\email{guillaume.schull@ipcms.unistra.fr}
\affiliation{Universit\'e de Strasbourg, CNRS, IPCMS, UMR 7504, F-67000 Strasbourg, France}

\author{Anna Ros\l awska}
\email{a.roslawska@fkf.mpg.de}
\affiliation{Universit\'e de Strasbourg, CNRS, IPCMS, UMR 7504, F-67000 Strasbourg, France}
\affiliation{Max Planck Institute for Solid State Research, Stuttgart, 70569, Germany}

\maketitle

\tableofcontents

\newpage
\noindent

\subsection*{S1 -- Methods}

The data are acquired with a low-temperature (6 K), UHV Unisoku USM1400 set-up with optical access and two adjustable lenses. A tunable laser source, composed of a supercontinuum laser (NKT Photonics SuperK Extreme EXW-6 or SuperK Fianium FIU-15) combined with an acousto-optic tunable filter (NKT Photonics SuperK SELECT) or a tunable high-resolution bandpass filter (NKT Photonics SuperK LLTF Contrast) provides narrow-band light ($<$ 2.5 nm) in the optical range of interest in this paper ($\approx$ 620 nm). This laser is treated as a tunable continuous light source\cite{Roslawska2023}. The laser output is directed towards a viewport of the scanning tunneling microscope (STM) and then focused onto the tip apex by one of the lenses. The light emitted or scattered from the tip-sample junction is collected by the second lens, filtered to remove the excitation line, and directed towards a single-pass monochromator coupled to a liquid nitrogen-cooled charge-coupled device array (Princeton Instruments). More details about the experimental set-up can be found in our earlier work \cite{Roslawska2023}.

\newpage

\subsection*{S2 -- Many-body model of the system}

\begin{figure}[!ht]
    \includegraphics[width = 0.4\textwidth]{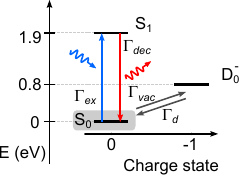}
    \caption{Schematics of the many-body transitions included in the model.}
    \label{SI_model}
\end{figure}

The dynamics of the single-molecule charging and discharging, responsible for the observed photoluminescence quenching, can be described by a straightforward many-body model using rate equations. The sketch of the relevant transitions is presented in \Figref{SI_model}. We include the excitation rate ($\Gamma_{ex}$), radiative decay rate ($\Gamma_{dec}$), charging rate ($\Gamma_{vac}$) and discharging rate ($\Gamma_{d}$). Note that for simplicity we do not consider the degeneracy of the states and orbitals in the model. These rates determine the time evolution of the populations of the ground state ($N_{S_0}$), excited state ($N_{S_1}$), and the charged state ($N_{D_0}$), which can be described by the following master equation:

\begin{equation}
\frac{\mathrm{d} }{\mathrm{d} t}\begin{bmatrix}
N_{S_0}\\
N_{S_1}\\
N_{D_0}
\end{bmatrix}=\begin{bmatrix}
-(\Gamma_{ex}+\Gamma_{vac}) & \Gamma_{dec} & \Gamma_{d} \\
\Gamma_{ex} & -\Gamma_{dec} & 0 \\
\Gamma_{vac} & 0 & -\Gamma_{d} \\
\end{bmatrix}\begin{bmatrix}
N_{S_0}\\
N_{S_1}\\
N_{D_0}
\end{bmatrix}
\end{equation}
We solve this system assuming that $\Gamma_{dec}$ is the fastest rate, since the exciton lifetime is estimated to be in the order of 1 ps \cite{Roslawska2022}, while the charging and discharging occur on the scale of 1 ns and the optical excitation is even less frequent, on the scale of 1 $\mu$s due to the low power used \cite{Roslawska2023}. Extracting the steady-state ($t  \xrightarrow{} \infty$) populations and profiting from $\Gamma_{dec}$ being the dominating term, we obtain:
\begin{equation}
    N_{S_0} = \frac{\Gamma_{d}\Gamma_{dec}}{\Gamma_{vac}\Gamma_{dec}+\Gamma_{d}\Gamma_{dec}+\Gamma_{d}\Gamma_{ex}} =  \frac{\Gamma_{d}}{\Gamma_{vac}+\Gamma_{d}+\frac{\Gamma_{d}\Gamma_{ex}}{\Gamma_{dec}} } = \frac{\Gamma_{d}}{\Gamma_{vac}+\Gamma_{d}}
    \label{NS0}
\end{equation}
\begin{equation}
    N_{S_1} = \frac{\Gamma_{d}\Gamma_{ex}}{\Gamma_{vac}\Gamma_{dec}+\Gamma_{d}\Gamma_{dec}+\Gamma_{d}\Gamma_{ex}} \approx 0
\end{equation}
\begin{equation}
    N_{D_0} = \frac{\Gamma_{vac}\Gamma_{dec}}{\Gamma_{vac}\Gamma_{dec}+\Gamma_{d}\Gamma_{dec}+\Gamma_{d}\Gamma_{ex}} = \frac{\Gamma_{vac}}{\Gamma_{vac}+\Gamma_{d}+\frac{\Gamma_{d}\Gamma_{ex}}{\Gamma_{dec}} }  = \frac{\Gamma_{vac}}{\Gamma_{vac}+\Gamma_{d}}
\end{equation}

For a molecule in a double-barrier (vacuum and NaCl) tunnel junction the charging rate through the vacuum depends on the tip-molecule distance:  $\Gamma_{vac}(\Delta z) = \Gamma_{vac}^0 \rm{exp(-2\kappa_{vac} \Delta z)}$ with the tunneling rate from the tip for $\Delta z$ = 0 $\Gamma_{vac}^{0}$ and the decay constant $\kappa_{vac}$, and $\Gamma_d=\frac{1}{\tau_d}$, where $\tau_d$ is the lifetime of the charged state. The first term describes the tunneling through the vacuum, which scales exponentially with the tip-molecule distance and the second term describes the neutralization via tunneling through a thin NaCl film \cite{Kaiser2023a}. Using these expressions in Eq.\eqref{NS0} we arrive at Eq. (2) in the main text.

We also note that:
\begin{equation}
    N_{S_1} = N_{S_0} \frac{\Gamma_{ex}}{\Gamma_{dec}}
\end{equation}
Therefore, the light intensity, usually described as the product of the excited state population and the radiative decay rate, can be written as $P(\Delta z) = \Gamma_{dec} N_{S_1} =  \Gamma_{ex} N_{S_0}$.
For completeness, we list the fit results in Table S1, with parameters as defined in the main text.

\begin{table}
\centering
\caption{Parameters fitted using the rate-equation model}
\vspace{0.3cm}
\begin{tabular}{c|c}
Parameter & Value \\
\hline
$\Gamma_{ex}^0$ & 5 $\times 10^3$ Hz \\
$\alpha$ & 4 $\times 10^9$ m$^{-1}$ \\
$\tau_{d}$ & 4 $\times 10^{-10}$ s\\
$\Gamma_{vac}^0$ & 4 $\times 10^{7}$ s$^{-1}$ \\
$\kappa_{vac}$ & 9 $\times 10^{9}$ m$^{-1}$ \\
\end{tabular}\label{stab:fit}
\end{table}
\newpage

\subsection*{S3 -- Photoluminescence quenching of H$_2$Pc}

\begin{figure}[!ht]
    \includegraphics{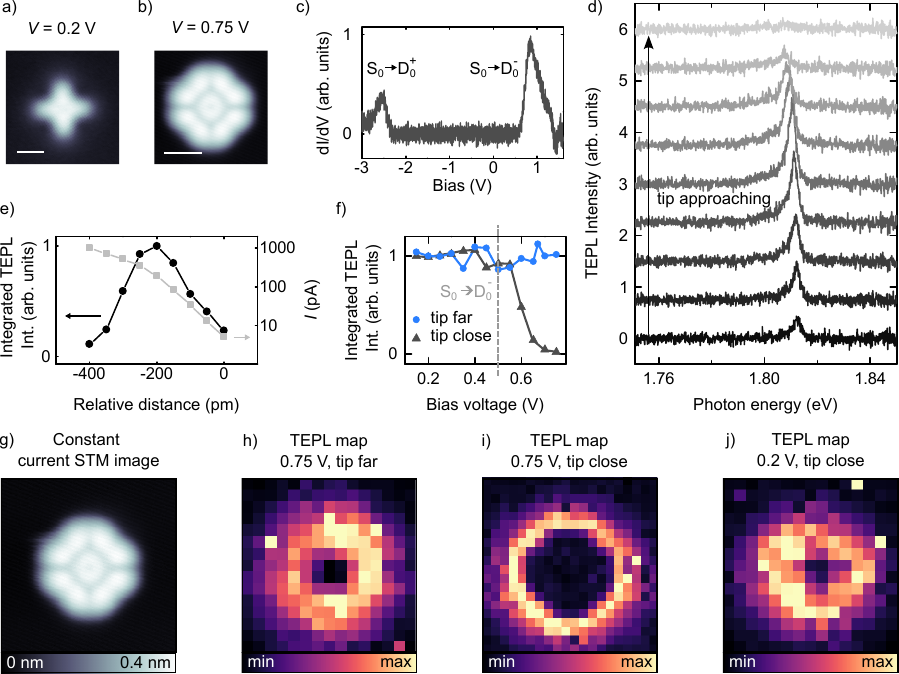}
    \caption{Photoluminescence quenching of H$_2$Pc on 3 ML NaCl/Ag(111). a,b) Constant current STM images of H$_2$Pc , $I$ = 2 pA (a) and $I$ = 5 pA (b), scale bar: 1 nm. c) d$I$/d$V$ spectrum of H$_2$Pc, set-point: $V$ = 1.6 V $I$ = 80 pA. d) TEPL spectra recorded on H$_2$Pc as a function of the relative tip-molecule distance, $P$ = 1 $\mu$W, $V$ = 0.75 V, $h\nu_{in}$ = 2 eV. The bottom spectrum is recorded at $I$ = 5 pA, then the tip is approached in 50 pm intervals. The traces are offset for clarity. e) Integrated TEPL intensity (range: 1.80 eV $<$ $h\nu_{out}$ $<$ 1.82 eV) from panel d), the relative distance is defined with respect to the initial conditions ($\Delta z$ = 0 for $V$ = 0.75 V, $I$ = 5 pA). Right axis: $I(\Delta z)$ measurement, $V$ = 0.75 V. f) TEPL intensity vs. bias at a constant tip-molecule distance. The blue circles are recorded for a relatively large tip-molecule distance (defined for $V$ = 0.75 V, $I$ = 30 pA), the grey triangles are recorded in the saturation regime (defined for $V$ = 0.75 V, $I$ = 730 pA), $P$ = 2 $\mu$W, $h\nu_{in}$ = 2 eV. g) Constant current STM image, $V$ = 0.75 V, $I$ = 5 pA. h-j) TEPL maps of H$_2$Pc recorded at $h\nu_{in}$ = 2 eV and $P$ = 1 $\mu$W (h) or $P$ = 3.5 $\mu$W (i, j). The maps are recorded at constant height conditions. Set-point : $V$ = 0.75 V, $I$ = 30 pA (h) or $V$ = 0.75 V, $I$ = 730 pA (i, j). The size of all maps in g-j) is 4.4 $\times$ 4.4 nm$^2$.   }
    \label{SI_H2Pc}
\end{figure}

\noindent 

In \Figref{SI_H2Pc} we demonstrate the photoluminescence quenching on another molecular system, free-base phthalocyanine (H$_2$Pc) adsorbed on 3 ML NaCl/Ag(111). \Figref{SI_H2Pc}a shows the in-gap STM image of H$_2$Pc ($V$ = 0.2 V), whereas \Figref{SI_H2Pc}b displays the STM image recorded at negative ion resonance (NIR, $V$ = 0.75 V). These regimes are identified in the d$I$/d$V$ spectrum presented in \Figref{SI_H2Pc}c. Next, we investigate the evolution of tip-enhanced photoluminescence (TEPL) spectra recorded at NIR when the tip approaches the molecule (\Figref{SI_H2Pc}d). At large relative tip-molecule distances, $\Delta z > $ -200 pm (bottom traces), we observe an emission line at approximately $h\nu_{out}$ = 1.81 eV, which corresponds to the $Q_x$ transition of H$_2$Pc (decay of the lowest excited state) \cite{Murray2011}. At small relative tip-molecule distances, $\Delta z <$ -200 pm (top traces), this emission is quenched. This is illustrated also in \Figref{SI_H2Pc}e, where we plot the integrated photoluminescence intensity as a function of the relative distance to the molecule (black dots). In parallel, we measure the tunneling current (grey squares) and find that at $\Delta z$ $<$ -200 pm, the $I(\Delta z)$ curve deviates from the exponential dependence (note the logarithmic scale), which marks the transition to the saturation regime \cite{Kaiser2023a}. Note that the saturation current is around 1 nA for H$_2$Pc on 3 ML NaCl/Ag(111), nearly an order of magnitude higher than for ZnPc on 4 ML. Similar to the discussion presented in the main text, the photoluminescence quenching occurs when the population of the $S_0$ state is reduced in favor of the population of the $D_0^-$ state. This is also demonstrated by the TEPL vs. bias dependence recorded at both low currents and in the saturation regime (\Figref{SI_H2Pc}f). When the tip is relatively far (blue circles), the TEPL intensity is insensitive to the bias voltage. However, when the tip is closer to the molecule (grey triangles), once the bias voltage is high enough for the $S_0$ $\xrightarrow{}$ $D_0^-$ transition to occur (marked by the grey dashed line), the $D_0^-$ state population increases significantly and the photoluminescence is quenched. The population of the $S_0$ state is also controlled by the tip position. Comparing the photoluminescence map at small $\Delta z$ (\Figref{SI_H2Pc}h) with the TEPL map recorded in the saturation regime at the same bias voltage (\Figref{SI_H2Pc}i), we find a broad low-intensity region in the central part of the latter map. This characteristic is absent at the map recorded at the same $\Delta z$ but at a bias voltage below the threshold for the $S_0$ $\xrightarrow{}$ $D_0^-$ transition (\Figref{SI_H2Pc}j). This effect is related to the fact that the transient charging occurs only when the charges are injected directly into the molecule from the STM tip and as such is limited by the spatial extent of the molecular orbital (\Figref{SI_H2Pc}g). We expect this effect to be general in TEPL measurements performed with a conductive tip on molecules adsorbed on insulating films.

\newpage

\bibliography{chargingTEPL}